# Femtosecond laser inscription of Bragg grating waveguides in bulk diamond

V. Bharadwaj[1,2,3], A. Courvoisier[4], T. T. Fernandez[1,2], R. Ramponi[1,2], G. Galzerano[1,2], J. Nunn[5], M. J. Booth[4], R. Osellame[1,2], S. M. Eaton[1,2,3], P. S. Salter[4,*]

[1]Istituto di Fotonica e Nanotecnologie-Consiglio Nazionale delle Ricerche (IFN-CNR), Piazza Leonardo da Vinci 32, Milano, Italy
[2]Department of Physics, Politecnico di Milano, Piazza Leonardo da Vinci 32, Milano, Italy
[3]Center for Nano Science and Technology, Istituto Italiano di Tecnologia, Milano, Italy
[4]Department of Engineering Science, University of Oxford, Parks Road, Oxford, UK
[5]Centre for Photonics and Photonic Materials, Department of Physics, University of Bath, North Road, Bath BA2 7AY, UK

*Corresponding author: patrick.salter@eng.ox.ac.uk

**Femtosecond laser writing is applied to form Bragg grating waveguides in the diamond bulk. Type II waveguides are integrated with a single pulse point-by-point periodic laser modification positioned towards the edge of the waveguide core. These photonic devices, operating in the telecommunications band, allow for simultaneous optical waveguiding and narrowband reflection from a 4th order grating. This fabrication technology opens the way towards advanced 3D photonic networks in diamond for a range of applications.**

Diamond has attracted great interest in the quantum optics community thanks to its nitrogen vacancy (NV) center, a naturally occurring impurity that is responsible for the pink coloration of diamond crystals. The NV spin state with the brighter luminescence yield can be exploited for spin readout, exhibiting millisecond spin coherence times at ambient temperature [1], comparable to trapped ions. In addition, the energy levels of the ground state triplet of the NV are sensitive to external fields. These properties make NVs attractive as a scalable platform for efficient nanoscale resolution sensing based on electron spins [2] and for quantum information systems [3]. In addition, the strong Raman coefficient of diamond is effective for Raman lasers [4]. These applications would benefit from a photonic platform, but due to diamond's hardness and chemical inertness, a mature fabrication toolkit has yet to be developed [5].

Recently, femtosecond laser writing demonstrated the formation of buried optical waveguides in single-crystal synthetic diamond [6, 7]. By laser inscribing two closely spaced modifications of reduced refractive index, 3D optical waveguides in bulk diamond were shown. Compared to reactive ion etching/photolithography [8, 9, 10] and ion irradiation [11] fabrication methods, femtosecond laser writing is advantageous due to its ability to pattern optical circuits out-of-plane with arbitrary 3D designs in a single process step. Femtosecond laser writing can also produce high quality single NVs on-demand [12] and even coupled to pre-existing laser written optical circuits [13].

In addition to 3D optical waveguides, narrowband reflection elements which permit wavelength-selective filtering and feedback would be of great benefit for optical magnetometry, quantum information and Raman laser applications. By adopting the Bragg grating waveguide (BGW) femtosecond laser fabrication method developed for glasses [14, 15], we demonstrate the first BGW in bulk diamond, for simultaneous optical waveguiding and narrowband reflection.

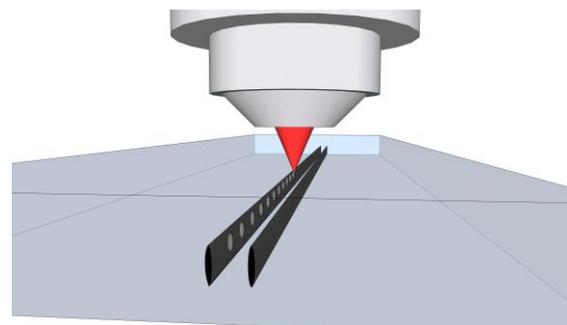

Fig. 1. Schematic Type II waveguide fabrication with embedded point-by-point grating shifted vertically upwards with respect to the waveguide.

Buried modifications were laser written in CVD single crystal diamond using a Ti:Sapphire laser (wavelength 790 nm, pulse duration 250 fs, maximum repetition rate 1 kHz). The diamond substrate was a 3 × 3 × 0.5 mm³ slab (Element 6, Optical grade) with nitrogen impurity levels < 1 ppm and absorption coefficient <0.1 cm⁻¹ at optical wavelengths. The crystal was cut with {100} faces, all of which were polished. A liquid crystal spatial light modulator (SLM) (Hamamatsu, X10468-02) was employed to correct for spherical aberration introduced during focusing due to refraction at the diamond interface [16]. The SLM was imaged onto the pupil plane of a high numerical aperture (NA) microscope objective (Olympus PlanApo 60×, 1.4 NA) using a 4$f$ system. Correction of spherical aberration enabled the formation of submicron-dimension buried modifications at various depths within the diamond sample. Further details of the experimental setup can be found in Ref. [17]. To translate the sample with respect to the focused laser beam, three-axis air bearing translation stages (ABL1000, Aerotech) were employed.

Type II waveguides were formed with 1 kHz repetition rate, 80 nJ pulse energy and 0.1 mm/s scan speed. Using the adaptive optics described above to compensate for aberration, nearly symmetric tracks were formed, with transverse and vertical dimensions of 1 μm and 2 μm, respectively. The side walls of the type II waveguide were built bottom up using 6 passes of the diamond through the laser focus, with a vertical translation of 3 μm per pass, leading to a final vertical dimension of 18 μm. For single mode waveguiding at the 1550-nm wavelength targeted in this study, a 25 μm separation between the side walls was employed.

There are two interesting consequences of the axial multi-pass fabrication strategy used to create waveguides in the diamond. Firstly, the deeper tracks, which are fabricated first, absorb more light on subsequent passes, such that the final side walls are broader than a single laser modification generating a greater strain. Equally these same tracks can seed further transverse fabrication during subsequent passes through the laser such that waveguides can be written right up to the sample edge [7], despite the strong optical aberrations present. Thus, after waveguide fabrication there was no need to re-polish the end-facet of the diamond prior to waveguide characterization.

After inscription of the side walls, a point-by-point periodic modification was written all along the waveguide by lowering the repetition rate to $R$ = 100 Hz and using a scan speed of $v$ = 0.13 mm/s. This resulted in a grating pitch of $\Lambda = v/R$ = 1.3 μm, as shown in Fig. 2a. The periodic modulation was written with 1.5 μJ pulse energy, ~20-fold higher than that used to form the side walls, in order to maximize the strain generated by the single pulse structural modifications and create greater overlap between the periodic refractive index and the waveguide core. We note that as a side effect of the raised pulse energy, the periodic element was significantly vertically elongated due to nonlinear propagation (Fig. 2b). The center of the vertically extended side walls is 50 μm below the surface of diamond and the bottom edge of the transversely-centered periodic modulation extends 3 μm below the top edge of the side walls, as shown in Fig. 2b. Following a recent study by Calmano *et al.,* the periodic modulation was shifted upwards with respect to the waveguide mode to minimize propagation loss, while also providing a strong Bragg resonance [18]. We note that when the periodic Type II modifications were fully contained within the waveguide core, no notable Bragg peak was observed.

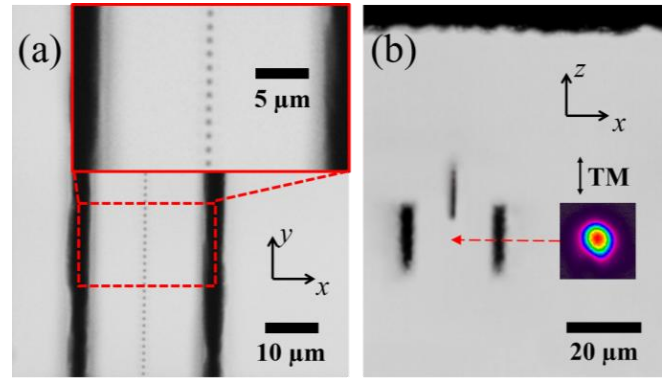

Fig. 2. (a) Overhead optical microscope image of type II waveguide with a point-by-point periodic modulation (1.3 μm pitch) in the center for Bragg reflection. (b) Optical microscope image of the end facet showing the type II waveguide with side walls separated by 25 μm. The bottom edge of the transversely-centered point-by-point periodic modulation extends 3 μm below the top edge of the side walls. Inset shows the near field intensity of the TM guided mode at 1560-nm wavelength.

The resonant wavelength is given by the well-known Bragg condition [19]:

$$\lambda_B = 2n_{\text{eff}}\Lambda/m \quad (1)$$

where $\lambda_B$ is the Bragg wavelength, $n_{\text{eff}}$ is the effective modal index (DC refractive index) and $m$ is the grating order. Estimating $n_{\text{eff}}$ as the bulk refractive index (2.3878 at 1550 nm [20]), a Bragg reflection at 1550 nm wavelength for a 4th order grating is predicted.

Using a standard fiber-waveguide characterization setup [21] the mode field diameter (MFD) of the BGW was found to be 10 μm × 10 μm (Fig. 2b inset) at 1560 nm wavelength, outside the Bragg reflection bandwidth. The single mode waveguides supported only the TM polarization (as indicated in Fig. 2b) and all subsequent results refer to the TM polarization state launched from polarization maintaining fiber. The out of band insertion loss of the BGW was 7.5 dB at 1560 nm, of which 3 dB can be directly attributed to the presence of the point by point grating.

Using a broadband amplified spontaneous emission (ASE) source (ASE-100-C, IPG Laser GmbH) together with an optical spectrum analyzer (AQ6317C, Ando) with 10 pm resolution, the spectral transmission properties of the BGW were characterized. As shown in Fig. 3a, a narrowband (FWHM = 290 pm) transmission dip was observed, centered at $\lambda_B$ = 1552.13 nm. The strength of the dip was 6.3 dB, impressively high for a 4th order grating. The ~1 dB loss on the short wavelength side of the resonance is attributed to radiation mode losses.

The maximum reflectivity of a Bragg grating is given by [22]:

$$R_{\max} = 1 - 10^{-\frac{T}{10}} = \tanh^2(\kappa L) \quad (2)$$

where $T$ is the transmission dip in dB, $\kappa$ is the coupling coefficient and $L$ is the grating length (3 mm). Solving for the coupling coefficient, we obtain $\kappa$ = 450 m$^{-1}$, yielding a grating strength $\kappa L$ = 1.4, of similar strength as BGWs laser-written in other crystalline media [23, 24, 18].

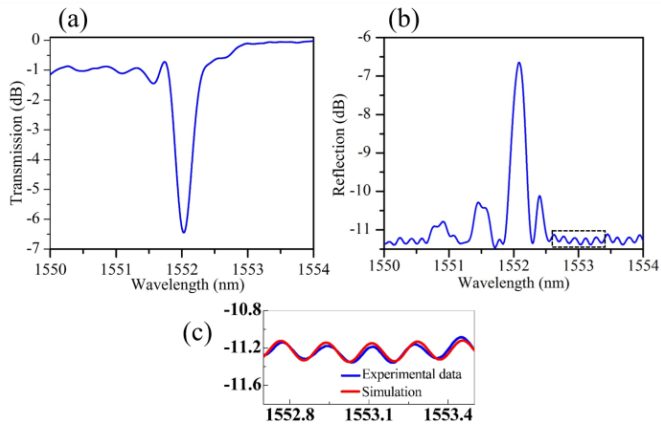

Fig. 3. Transmission (a) and reflection (b) spectra of Bragg grating waveguide in diamond. (c) shows simulation of the Fabry-Perot fringes within the dashed region in (b) used to calibrate the reflection spectrum.

Figure 3b shows the recorded reflectivity spectrum. For glass-based BGWs, the reflection spectrum may be calibrated using a known reflection from a fiber Bragg grating butt-coupled to the output facet of the waveguide [15]. To calibrate the reflectivity peak of the grating here, we exploited the out-of-band Fabry-Perot (FP) fringes due to the Fresnel reflections at the diamond interfaces. An index matching gel ($n_{gel}$ =1.46) was used to avoid air-diamond/air-fiber interfaces and reduce the Fresnel reflection coefficient from 17% to approximately 6% ($n_{diamond}$ = 2.38 and $n_{fiber}$ = $n_{gel}$ = 1.46). A numerical simulation [25] was used to provide detailed analysis of the observed fringes as shown in Fig. 3c, estimating a propagation loss of 4.6 dB for the 3 mm long diamond Bragg waveguide. The corresponding value of the Bragg reflection peak is -6.5 dB, 5 dB higher than the FP fringe background. From the full width at first zeroes bandwidth of the reflection spectrum [21]:

$$\Delta\lambda_{\text{FWFZ}} = \frac{\lambda^2}{\pi n_{\text{eff}} L}\sqrt{(\kappa L)^2 + \pi^2} = 367 \text{ pm} \quad (3)$$

we deduce a coupling coefficient of $\kappa$ = 460 m$^{-1}$, in excellent agreement with the transmission spectrum and Eq. (2).

In summary, we have demonstrated laser written waveguides in the diamond bulk integrated with fourth order point by point Bragg gratings to give narrowband reflection in the telecommunications band. In future experiments, we will seek to form finer pitched modulations within the type II diamond waveguides (< 500 nm), targeting shorter wavelength Bragg resonances for use in applications such as Raman lasers [4] and magnetometry [26, 27]. Alternatively, taking advantage of diamond's ultrawide transparency window deep in the THz, larger pitched modulations will also be pursued to enable Bragg grating waveguides for a versatile range of sensing applications at mid to far infrared wavelengths [28].

**Funding.** SME is thankful for support from the DIAMANTE MIUR-SIR grant. PSS and MJB gratefully acknowledge the Leverhulme Trust (RPG-2013-044) and the UK Engineering and Physical Sciences Research Council (EP/K034480/1) for financial support.
**Acknowledgements.** We thank Dr. Haibin Zhang, Dr. J. P. Hadden and Prof. Michael Withford for enlightening scientific discussions.